\documentstyle[aps,prl,epsfig]{revtex}
\tightenlines
\begin{document}
\twocolumn[\hsize\textwidth\columnwidth\hsize\csname@twocolumnfalse\endcsname
\draft
\title{Interacting Growth Walk - a model for hyperquenched homopolymer glass?}
\author{S.L. Narasimhan$^{*}$,  P.S.R. Krishna, A. K. Rajarajan
 and K.P.N. Murthy$^{\dagger}, {}^{\dagger\dagger}$}
\address{Solid State Physics Division,  Bhabha Atomic Research Centre,
\\ Mumbai - 400 085, India}
\address{$^{\dagger}$ Institut f\"ur Festk\"orperforsuchung, 
Forschungszentrum J\"ulich GmbH,\\ D-52425 J\"ulich, Germany}
\maketitle

\begin{abstract}
We show that the compact self avoiding walk configurations, kinetically
generated by the recently introduced Interacting Growth Walk (IGW) model,
can be considered as members of a canonical ensemble if they are assigned
random values of energy. Such a mapping is necessary for studying the
thermodynamic behaviour of this system. We have presented the specific heat
data for the IGW, obtained from extensive simulations on a square lattice; we
observe a broad hump in the specific heat above the $\theta$-point, contrary
to expectation.
\end{abstract}
\vfill
\twocolumn
\vskip.5pc]
\narrowtext

Linear polymers in a poor solvent are known [1] to assume globular
configurations below a tricritical temperature $T_{\theta}$,
called the $\theta$-point. These globules acquire denser
minimum energy configurations at lower temperatures. In the case
of random heteropolymers, the 'quenched' random interactions
between the constituent monomers frustrate the evolution of the
globules towards their minimum energy configurations. They are thus
forced to freeze into higher energy configurations (local
minima). In fact, the heteropolymer globules serve as 'toy models'
for protein folding phenomenon [2]. It has been shown recently [3]
that even homopolymer globules can freeze into glassy states, due to
a self-generated disorder brought about by the competing interactions
and chain connectivity during the cooling process. In this sense,
the freezing of a homopolymer globule is said to be analogous to that
of a structural glass.

In a Monte Carlo study of this freezing process, we may choose a
configuration from a canonical ensemble of Interacting Self Avoiding
Walks (ISAW) [4] which represents a linear polymer in equilibrium with a
thermal bath at a temperature $T$ (say, $\geq T_{\theta}$). Then,
using a standard dynamical algorithm [5], we may relax the chosen
configuration at a temperature preset ({\it i.e.,} quenched) to a
desired value less than $T_{\theta}$; deeper the quench, more
difficult and time consuming it would be to realize a globular
configuration. On the other hand, the
Interacting Growth Walk (IGW) [6] is a simpler but more efficient
algorithm for generating compact or globular Self Avoiding Walks
(SAW); they are generated, step by step, by sampling the locally
available sites with appropriate Boltzmann factors,
$exp(\beta _{G} n_{NN}^{m}\epsilon _{0})$, where $\beta _{G}^{-1}$ is the
'growth' temperature, $n_{NN}^{m} (1\leq m\leq z-1)$ is the number of
non-bonded nearest neighbour (nbNN) contacts the site $m$ will make,
if chosen, on a lattice of coordination number $z$ and $-\epsilon _{0}$
is the attractive energy associated with any nbNN contact.

In this paper, we show that these kinetically generated IGWs
represent the frozen configurations of a homopolymer globule with
a self-generated disorder. Contrary to expectation, our
simulations on a square lattice indicate an excess specific heat,
characterizing these frozen states, above the $\theta$-point. In fact, 
this simple model demonstrates that a meaningful statistical mechanical
description of an irreversible growth process involves an element of
self-generated disorder brought about by ergodicity-breaking of the system.

The growth of an IGW starts by first "occupying" an arbitrarily
chosen site ${\bf r}_{0}$ of a regular $d$-dimensional lattice of
coordination number $z$ whose sites are initially "unoccupied"
(by monomers). The first step of the walk is taken in one of the
$z$ available directions by choosing an "unoccupied" nearest
neighbours (NN) of ${\bf r}_{0}$, say ${\bf r}_{1}$, at random and
with equal probability. Let the walk be nonreversing so that it has
a maximum of $z-1$ directions to choose from for the next step.
Let $\{ {\bf r}_{j}^{m} \mid m = 1,2,...,z_{j}\}$ be the "unoccupied"
NN's available for the $j$th step of the walk. If $z_{j} = 0$, the
walk cannot grow further because it is geometrically "trapped".
It is, therefore, discarded and a fresh walk is started from
${\bf r}_{0}$. If $z_{j} \not= 0$, the walk proceeds as follows:

Let $n_{NN}^{m}(j)$ be the number of nbNN sites of ${\bf r}_{j}^{m}$.
Then, the probability that this site is chosen for the $j$th step
is given by,
\begin{equation}
p_{m}({\bf r}_{j}) \equiv \frac{exp[\beta _{G}n_{NN}^{m}(j)\epsilon _{0}]}
                               {\sum _{m=1}^{z_{j}} exp[\beta _{G}n_{NN}^{m}(j)\epsilon _{0}]}
\end{equation}
where the summation is over all the $z_{j}$ available sites. At
"infinite" temperature ($\beta _{G} =0$), the local growth probability
$p_{m}({\bf r}_{j})$ is equal to $1/z_{j}$ and thus, the walk
generated will be the same as the Kinetic Growth Walk (KGW)[7].
However, at finite temperatures, the walk will prefer to step into
a site with more nbNN contacts. We have illustrated this local growth
rule in Fig.1(a) for IGW on a square lattice. Lower the growth
temperature, less is the attrition (see the inset of Fig.2) that the walk
suffers while also being able to grow into more compact configurations.
Moreover, it has been shown [6] that a $\theta$-point for this walk exists,
and that the walk belongs to the same universality class (${\it i.e.,}$ has
the same values of the universal exponents, $\nu$ and $\gamma$) as
the SAW above, at and below the $\theta$-point.

%%%%%%%%%%%%%%%%%%%%%%%%%%%%%%%%%%%%%%%%%%%%%%%%%
%                FIG. 1
%%%%%%%%%%%%%%%%%%%%%%%%%%%%%%%%%%%%%%%%%%%%%%%%%
%\newpage
\begin{figure}
\setlength{\unitlength}{0.25in}
\begin{picture}(13,11)
\linethickness{0.075mm}
%\multiput(1,1)(1,0){12}{\line(0,1){11}}
%\multiput(1,1)(0,1){12}{\line(1,0){11}}
\linethickness{0.25 mm}
\put(0,7){\circle{0.35}}

\put(0,7){\vector(1,0){0.9}}
\put(1,7){\circle*{0.35}}

\put(1,7){\vector(0,1){0.9}}
\put(1,8){\circle*{0.35}}

\put(1,8){\vector(1,0){0.9}}
\put(2,8){\circle*{0.35}}

\put(2,8){\vector(0,1){0.9}}
\put(2,9){\circle*{0.35}}

\put(0.875,9){$\times$}
\put(0.5,9){A}

\put(1.875,10){$\times$}
\put(2,10.5){B}

\put(2.875,9){$\times$}
\put(3.5,9){C}

\put(1,6){(a)}

\put(10.5,7){\circle{0.35}}

\put(10.5,7){\vector(0,1){0.9}}
\put(10.5,8){\circle*{0.35}}

\put(10.5,8){\vector(0,1){0.9}}
\put(10.5,9){\circle*{0.35}}

\put(10.5,9){\vector(0,1){0.9}}
\put(10.5,10){\circle*{0.35}}

\put(10.5,10){\vector(1,0){0.9}}
\put(11.5,10){\circle*{0.35}}

\put(11.5,10){\vector(1,0){0.9}}
\put(12.5,10){\circle*{0.35}}

\put(12.5,10){\vector(0,-1){0.9}}
\put(12.5,9){\circle*{0.35}}

\put(12.5,9){\vector(-1,0){0.9}}
\put(11.5,9){\circle*{0.35}}

\put(11.5,9){\vector(0,-1){0.9}}
\put(11.5,8){\circle*{0.35}}

\put(11.375,7){$\times$}
\put(11.75,6.75){A}

\put(12.375,8){$\times$}
\put(13,8){B}

\put(12.5,6){(c)}

\put(7,8){\circle{0.35}}

\put(7,8){\vector(0,1){0.9}}
\put(7,9){\circle*{0.35}}

\put(7,9){\vector(1,0){0.9}}
\put(8,9){\circle*{0.35}}

\put(8,9){\vector(0,1){0.9}}
\put(8,10){\circle*{0.35}}

\put(8,10){\vector(-1,0){0.9}}
\put(7,10){\circle*{0.35}}

\put(7,10){\vector(-1,0){0.9}}
\put(6,10){\circle*{0.35}}

\put(6,10){\vector(0,-1){0.9}}
\put(6,9){\circle*{0.35}}

\put(6,9){\vector(0,-1){0.9}}
\put(6,8){\circle*{0.35}}

\put(6,8){\vector(0,-1){0.9}}
\put(6,7){\circle*{0.35}}

\put(6.875,6.875){$\times$}
\put(7,6.25){A}

\put(5.875,5.875){$\times$}
\put(6,5.25){B}

\put(4.865,6.875){$\times$}
\put(5,6.25){C}

\put(8,6){(b)}
\end{picture}
\vspace{-1.25in}
\caption{A simple illustration of the IGW algorithm for generating walks from the origin, denoted by the
open circle, at a given growth temperature, $\beta _{G}^{-1}$. (a) The sites A, B and C are available
for making the fifth step. Choosing the site A will lead to one nbNN contact, whereas choosing the
sites B or C will lead to none. Hence, the sites A, B and C will be chosen with probabilities
$e^{\beta_{G}}/(2+e^{\beta_{G}})$,
$1/(2+e^{\beta_{G}})$ and $1/(2+e^{\beta_{G}})$ respectively.
(b) The probability of growing this configuration is given by $p_{b} = (1/4)(1/3)^{2}(1/2)^{2}(e^{\beta_{G}}/[2+e^{\beta_{G}}])^{2}(e^{\beta_{G}}/[1+e^{\beta_{G}}])$.
(c) The probability of growing this configuration, which is identical
to (b), is given by $p_{c} = (1/4)(1/3)^{5}(e^{2\beta_{G}}/[2+e^{2\beta_{G}}])$.}
\label{walk}
\end{figure}
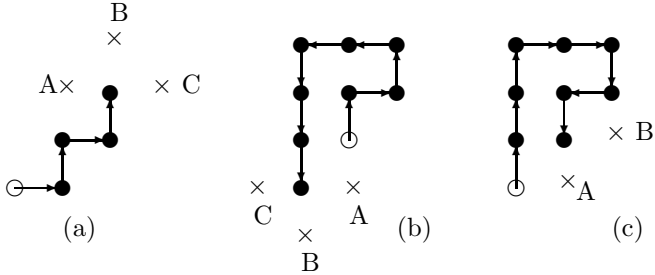
%%%%%%%%%%%%%%%%%%%%%%%%%%%%%%%%%%%%%%%%%%%%%%%%%

We have repeated the IGW simulations on a square lattice for walks
upto $N = 8000$, much longer than reported in ref.[6] and with better
statistics. In Fig.2, we have shown the $N$-dependence of the exponent,
$\nu (N)$, obtained from the mean squared radius of gyration data, for
various values of $\beta _{G}$ in the range $3$ to $10$. We have estimated
the asymptotic values of this exponent as simple polynomial extrapolations
of these $\nu (N)$ values, and presented them in Fig.3, along with also those
obtained for $\beta = 0, 1, 1.5$ and $2$ from the earlier data reported
in ref.[6].

The transition from the SAW phase ($\nu = 3/4$) to the collapsed walk
phase ($\nu = 1/2$) seems to be taking place over a narrow range of
$\beta _{G}$ values ($\sim 3.5 \leq \beta _{G}\leq \sim 5.0$), but this
could still be due to limitations of our numerical work.
The asymptotic estimates of $\nu$ could improve not only with longer
walks but also with larger number of successful walks, and this could
result in narrower transition regime. The $\theta$-point for the IGW
corresponds to a growth temperature given by $\beta _{G} \sim 4.5$, which
is close to our earlier value ($\sim 4$) [6]. Thus, we see that IGW has
all the three distinct phases (extended, $\theta$-point and collapsed)
of SAW, realizable by tuning the growth temperature $\beta _{G}^{-1}$.

However, the IGW does not represent a homopolymer in equilibrium with
its environment at some bath temperature. Because, the set of all $N$-step
IGWs generated at a given growth temperature, $Z_{IGW}(N;\beta _{G})$,
is not equivalent to the canonical ensemble of ISAWs, $Z_{ISAW}(N;\beta)$,
for some bath temperature $\beta ^{-1}$. For example, in Fig.1(b) and 1(c),
we have shown two identical configurations which are expected to occur
with the same probability in a canonical ensemble, but are in fact grown
with different probabilities. This is a consequence of the fact that the
local growth probability, $p_{j}({\bf r}_{j})$, of making the $j$th step
to a site ${\bf r}_{j}$ depends on all the previous sites visited. Hence,
the probability of generating an IGW configuration,
${\cal C} \equiv \{ {\bf r}_{0},{\bf r}_{1},...,{\bf r}_{j},...\}$,
has to be written as $P_{IGW}(N,{\cal C}) = \prod _{j=1}^{N}p_{j}({\bf r}_{j};
{\bf r}_{0},{\bf r}_{1},...,{\bf r}_{j-1})$. Nonetheless, there must be
a correspondence between the kinetically generated IGW and the canonical
ISAW, especially because the former can be tuned to belong to the same
universality classes as the latter.

%%%%%%%%%%%%%%%%%%%%%%%%%%%%%%%%%%%%%%%%%%%%%%%%%%%%%
%                       FIG. 2                                                                                                    %
\begin{figure}
\includegraphics[width=3.25in,height=2.25in]{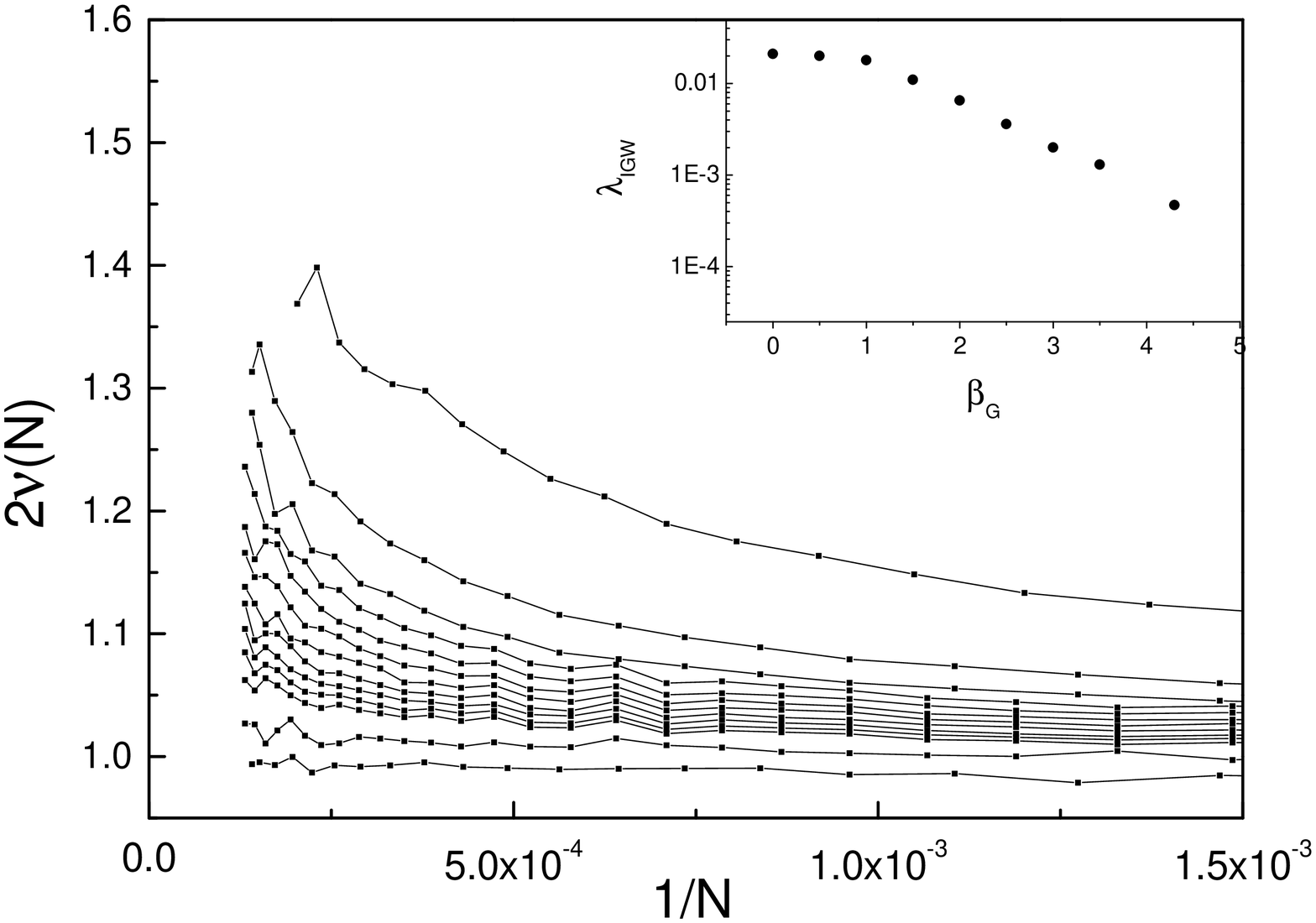}
\caption{The trend towards the asymptotic values of the exponent, $\nu$,
for various values of $\beta _{G}$s (=3.0, 3.5, 3.7, 3.8, 3.9, 4.0, 4.1, 4.2,
4.3, 4.4, 4.5, 5.0 and 10.0, from top to bottom. Inset:
Semi-logarithmic plot of the attrition constant as a function of $\beta _{G}$.
The data seem to suggest a form, $\lambda _{IGW} \propto exp(-a\beta _{G})$,
where $a$ is a constant.}

%\end{figure}
%%%%%%%%%%%%%%%%%%%%%%%%%%%%%%%%%%%%%%%%%%%%%%%%%%%%%
%%%%%%%%%%%%%%%%%%%%%%%%%%%%%%%%%%%%%%%%%%%%%%%%%%%%%
%                       FIG. 3                                                                                                    %
%\begin{figure}
\includegraphics[width=3.35in,height=2.3in]{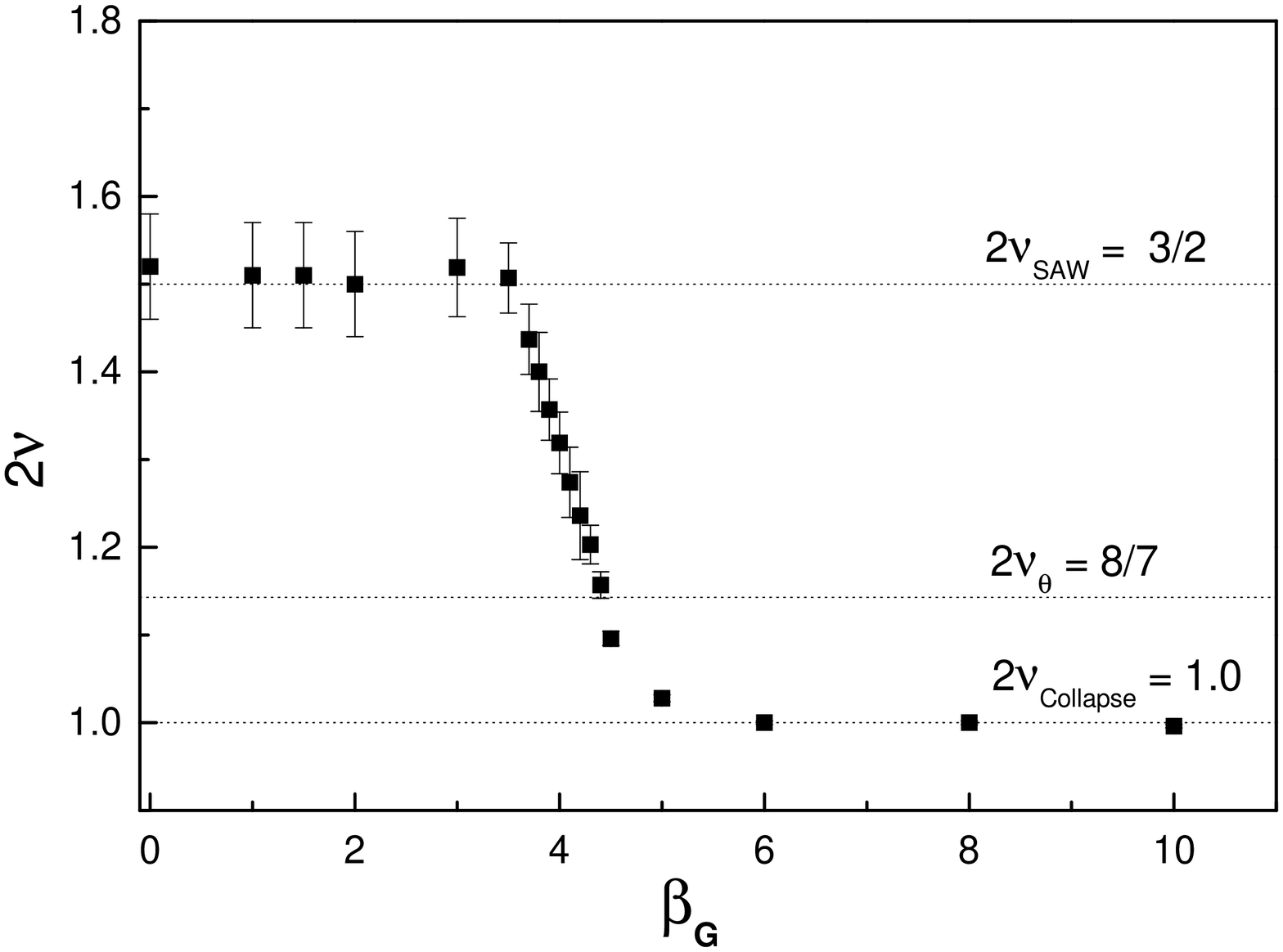}
\caption{The collapse scenario of IGW as brought out by the temperature dependence of $\nu$. }
\end{figure}
%%%%%%%%%%%%%%%%%%%%%%%%%%%%%%%%%%%%%%%%%%%%%%%%%%%%%
Let ${\cal E}_{G} \equiv \beta _{G}\epsilon _{0}$ denote the dimensionless
energy per nbNN contact at the growth tepmperature $\beta _{G}^{-1}$. Then,
an $N$-step IGW configuration, ${\cal C}$, having a total of $N_{c}({\cal C})$
such contacts will have an energy, $E_{G}({\cal C}) = {\cal E}_{G}N_{c}({\cal C})$.
As illustrated in Fig.1(b) and 1(c), configurations with the same energy are
generated with different probabilities. We may rewrite the growth probability,
$P_{IGW}(N,{\cal C})$, as follows.
\begin{eqnarray}
P_{IGW}(N;{\cal C}) & = & \prod _{j=1}^{N} p_{j}({\bf r}_{j};{\bf r}_{0},{\bf r}_{1},...,{\bf r}_{j-1})\\
%                    & = & \bigg(\prod _{j=2}^{N}\big[ (z-1)p_{j}({\bf r}_{j};{\bf r}_{0},{\bf r}_{1},...,{\bf r}_{j-1})\big] \bigg) \nonumber\\
%                     &  &\times  \bigg[\frac{1}{z} \bigg(\frac{1}{z-1}\bigg)^{N-1}\bigg]\\
%        & &\nonumber\\
                    & \equiv & e^{{\cal E}({\cal C})N_{c}({\cal C})}P_{SAW}(N)
\end{eqnarray}
where $P_{SAW}(N) \equiv z^{-1}(z-1)^{-(N-1)}$ is the probability of
generating an $N$-step SAW configuration and ${\cal E}({\cal C})$ is the energy
per contact to be assigned to the configuration if it were to be considered
as a member of a canonical ensemble.
\begin{equation}
{\cal E}({\cal C}) \equiv \frac{1}{N_{c}({\cal C})}
                   \sum _{j=2}^{N}\log \big[(z-1)p_{j}({\bf r}_{j};{\bf r}_{0},{\bf r}_{1},...,{\bf r}_{j-1})\big]
\end{equation}
It is now clear that different configurations with the same number contacts
could be assigned different values of ${\cal E}({\cal C})$ because their growth
probabilities are different. In other words, for a given value of the
growth parameter, ${\cal E}_{G}$, the mapping of IGW to ISAW gives rise to a
distribution of the dimensionless energy per contact, ${\cal E}$.

Assuming that $\epsilon _{0}$ is a constant, a distribution in ${\cal E}$
corresponds to a distribution in $\beta$. This implies that the IGW
configurations grown at a given temperature $\beta _{G}^{-1}$ can be
considered as ISAW configurations, but sampled at temperatures drawn from a
distribution in $\beta$. We have discussed this recently for IGW on a
honeycomb lattice [8]. We have shown that a sharply peaked distribution
in $\beta$ can be associated with any given $\beta _{G}>0$ (the broadest
distribution, numerically obtained for $\beta _{G} =\infty$, peaks at
$\beta \sim 1.21$ with a FWHM $\sim 0.03$). In the athermal limit
($\beta _{G} = 0$), the IGW corresponds to ISAW at a unique temperature
given by $\beta = \log 2$, a result obtained first by Poole {\it et al} [9].
Since the distribution in $\beta$ is sharp, the peak value may be taken to
provide a well defined canonical or 'bath' temperature at which most of the IGW
configurations can be considered as ISAW configurations. The ones that
correspond to different temperatures will have to be equilibrated
at the peak temperature.

Alternatively, if IGW were to be considered as an ISAW, then it should
represent an equilibrium configuration at a uniquely defined 'bath'
temperature. We fix the bath temperature, $\beta$, by assuming that the peak
position of the distribution in ${\cal E}$ can be identified with
$\beta \epsilon _{0}$. There is no {\it a priori} reason to assume that the
average energy per contact for the equilibrium configuration should be the
same as $\epsilon _{0}$, a parameter introduced for sampling the locally
available sites during its growth. Hence, the distribution in ${\cal E}$ can
be taken to be proportional to a distribution in $\epsilon$, peaking at
$\epsilon _{0}$.

%%%%%%%%%%%%%%%%%%%%%%%%%%%%%%%%%%%%%%%%%%%%%%%%%%
%                       FIG. 4
\begin{figure}
\includegraphics[width=3.25in,height=2.25in]{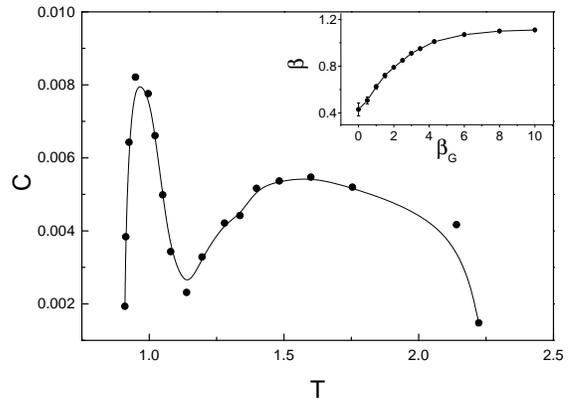}
\caption{Specific heat as a function of bath temperature,
$T \equiv \beta ^{-1}$. The sharp peak at $T \sim 1$ corresponds to
$\beta _{G} \sim 4.5$, and hence to the $\theta$-collapse transition.
The continuous line is a guide to the eye. Inset: Inverse of bath temperature,
$\beta $,
as a function of the inverse of growth temperature, $\beta _{G}$.}
\end{figure}
%%%%%%%%%%%%%%%%%%%%%%%%%%%%%%%%%%%%%%%%%%%%%%%%%%

We have obtained the bath temperature, $\beta (N)$, and the width, $\sigma (N)$,
of the distribution in $\epsilon$ as a function of $N$ for a given
$\beta _{G}$, basically from the first and second moments of the distribution
in ${\cal E}$. Then, we have estimated their asymptotic values by fitting
them to a simple form, $y(N) = y + (A/\ N^{B})$ where $y (=\beta$ or $\sigma)$,
$A$ and $B$ are adjustable parameters. We have presented the estimated $\beta$
values as a function of $\beta _{G}$ in the inset of Fig.4. We find that the full range of
$\beta _{G}\in [0,\infty]$ is mapped into a narrow range of bath temperatures,
$\beta \in [\sim 0.42,\sim 1.12]$ ($\in [\log 2,\sim 1.2]$, on honeycomb
lattice [8]). It may be noted that the $\theta$-point, $\beta _{G} \sim 4.5$,
corresponds to $\beta \sim 1$.

From the asymptotic variances, $\sigma ^{2}(\beta)$, we have obtained the specific
heat per contact, $c(\beta) = \beta ^{2}\sigma ^{2}(\beta)$, and presented them
in Fig.4 as a function of the bath temperature $\beta ^{-1}$. The sharp
peak seen at about $\beta \sim 1$ corresponds to the collapse transition
at the $\theta$-point. This, in fact, validates the view that a definite
bath temperature can be associated with the IGW.

But, there is no known transition that can be associated with the excess
specific heat seen as a broad hump above the $\theta$-peak, because this
region is in the SAW phase as far as the universal exponents are concerned (Fig.3).
It is therefore of interest to understand what is responsible for this excess
specific heat. Recently, hyperquenched glasses have been shown [10] to exhibit
excess specific heat (Fig.4 of Ref.[10]), strikingly similar to what we have
observed for the IGW (Fig.4) above the $\theta$-point. The dimensionless
energy per contact, ${\cal E}({\cal C})$, defined in Eqn.4, is indeed an
average of such values that can be evaluated during the growth process. This
implies that a distribution of ${\cal E}$ can be associated with every
configuration generated. Moreover, the IGW configurations are clearly much more
compact (see Fig.1 of ref.[6]) than the typical SAWs belonging to the same
universality class. It is therefore reasonable to consider them as "frozen"
globules.
%%%%%%%%%%%%%%%%%%%%%%%%%%%%%%%%%%%%%%%%%%%%%%%%%%
%                       FIG. 5
\begin{figure}
\includegraphics[width=3.4in,height=2.7in]{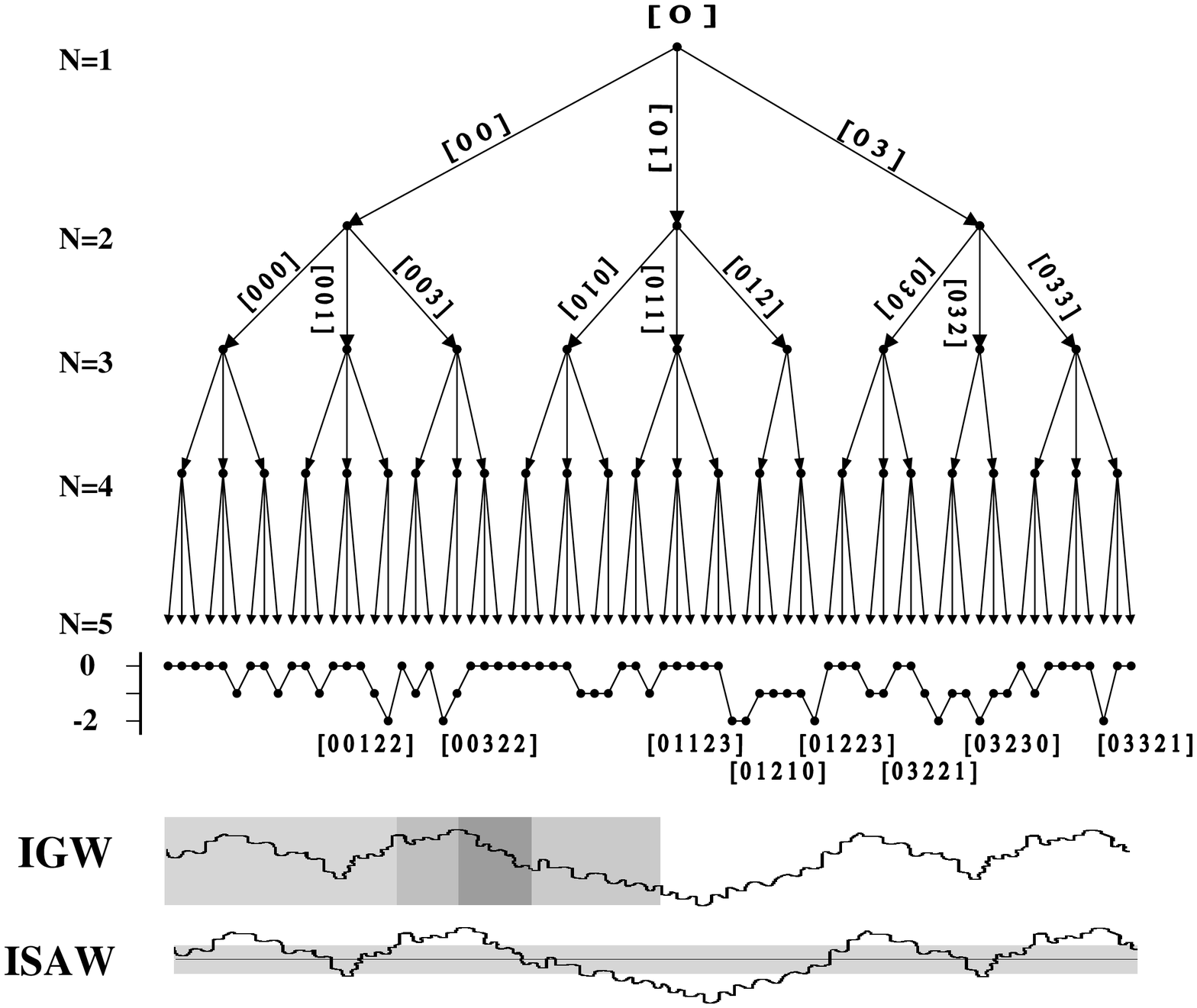}
\caption{A schematic illustration of how the growth of an IGW can be viewed
as a hierarchical process. The configurations are coded as strings of $0$s,
$1$s, $2$s and $3$s, enclosed within square brackets, where the labels
$0,1,2$ and $3$ correspond to steps in the $+x, -y, -x$ and $+y$ directions
respectively. The various paths in the hierarchy are taken with different
probabilities (see text). The tree is constructed in such a way that the
final configurations are numbered in increasing order from left to right.
Shown just below the tree is the energy landscape for all the $5$-step
walks whose first step is along the $+x$ direction. Of course, the
probability of realising a point on the landscape depends on the growth
temperature, $\beta _{G}^{-1}$. The global minimum energy ($=-2$)
configurations are indicated by their respective codes. And below this is
a schematic picture of the energy landscape for asymptotically long walks.
In the case of IGW, the number of available (or realisable) final configurations
decreases as the walk proceeds to grow. This is illustrated by shaded regions
becoming progressively darker. Exactly which point on the landscape is finally
reached is decided by the value of $\beta _{G}^{-1}$. In the case of ISAW,
however, all the configurations having their energies within an interval
(schematically indicated by the shaded region) determined by $\beta ^{-1}$
will be sampled.}
\end{figure}
%%%%%%%%%%%%%%%%%%%%%%%%%%%%%%%%%%%%%%%%%%%%%%%%%%

It may be noted that the correspondence between $\beta _{G}$ and $\beta$ whose
existence is dictated by Eqn.(4) forms the basis of this study. And, the fact
that the full range of $\beta _{G} \in [0,\infty]$ maps into a finite range of
canonical $\beta \in \ \sim [0.42,1.12]$ has subtle physical implications.
For example, as depicted in Fig.5, the growth of an IGW can also be considered
as a hierarchical approach towards realizing a particular configuration.
Every step taken reduces the number of available configurations, or equivalently,
restricts the accessible region of the energy landscape in a progressive
manner. This implies that irreversible growth is equivalent to breaking the ergodicity
of the system. The probability of taking a certain path in the hierarchy depends on
the tuning parameter, $\beta _{G}$. On the other hand, in the canonical
ensemble picture, we sample all the configurations whose energies lie within
an interval defined by the bath temperature, $\beta ^{-1}$ (schematically
illustrated in Fig.5). In particular, we expect to sample only those
configurations with global minimum energy when $\beta ^{-1} = 0$. In contrast,
with $\beta _{G}^{-1} = 0$, the IGW algorithm will generate a few zero energy
(athermal) configurations as well, besides those with global minimum energy;
hence, the corresponding $\beta ^{-1}$ will be greater than zero. And, larger
the value of the coordination number, $z$, of the lattice, smaller will be the
number of such athermal configurations and hence larger will be the value of
$\beta ^{-1}$ to which it corresponds. Similarly, the distribution of NN
contacts for the IGW configurations generated at $\beta _{G} = 0$ deviates
from that obtainable for SAW, and hence the corresponding $\beta$ will be
an $z$-dependent nonzero value.

In summary, we have shown that the IGW configurations can be considered
as members of a canonical ensemble ({\it i.e.,} as ISAW configurations) if
the energy per contact can be considered as a random variable. In general, 
a meaningful statistical mechanical description of an irreversible growth process 
involves an element of self-generated disorder. The signature
of this is seen as a broad hump in the specific heat above the $\theta$-
point. That these configurations are generated in an hierarchical manner,
as implied by the specific growth rule, provides additional support to the
conjecture that they may be taken to represent hyperquenched polymer
configurations. Conformational dynamics of IGW could throw further
light on this conjecture. In fact, the IGW seems to illustrate the generic
possibility of a growth process giving rise to hyperquenched states of a
system, if it is faster than the configurational relaxation.

SLN is grateful to R. Chidambaram and M. Rama\-nadham for inspiring him to
study the physics of growth walks. A part of the computational work was
carried out at the Institut f\"ur Festk\"orperforschung. K.P.N. thanks
Forschungszentrum J\"ulich for the hospitality extended to him during
March - April 2002. He also thanks V. Sridhar for fruitful discussions.
We thank P. V. S. L. Kalyani for help in preparing the figures.
\\
\noindent $^*$slnoo@magnum.barc.ernet.in;

\noindent ${}^{\dagger\dagger}$ Permenant address: 
Materials Science Division,\\
Indira Gandhi Centre for Atomic Research,\\ Kalpakkam 603 012, 
Tamilnadu, India.


\begin{references}
\bibitem{1} P. G. de Gennes, {\it Scaling Concepts in Polymer Physics}
                             (Cornell Univ. Press, NY,1979);
            C. Vanderzande, {\it Lattice models of polymers}
                             (Cambridge Univ. Press, Cambridge,1998).
\bibitem{2} H. S. Chan and K. A. Dill, Physics Today, {\bf 46}, 24 (1993);
            V. S. Pande, A. Yu. Grosberg and T. Tanaka, Rev. Mod. Phys. {\bf 72}, 259 (2000);
            {\it Protein Folding}, Edited by T. E. Creighton (Freeman, NY, 1992).
\bibitem{3} V. G. Rostiashvili, G. Migliorini and T. A. Vilgis,
            Phys. Rev. {\bf E64}, 051112 (2001);
            R. Du, A. Yu. Grosberg, T. Tanaka and M. Rubinstein,
            Phys. Rev. Lett. {\bf 84}, 2417 (2000); N.V. Dokholyan, E. Pitard,
            S.V. Buldyrev and H.E. Stanley, Phys. Rev. {\bf E65}, 030801(R)  (2002).
\bibitem{4} H. Saluer, J. Stat. Phys. {\bf 45}, 419 (1986);
            B. Duplantier and H. Saluer, Phys. Rev. Lett. {\bf 59}, 539 (1987);
            A. Baumgartner, J. Phys.(Paris), {\bf 43}, 1407 (1982);
            K. Kremer, A. Baumgartner and K. Binder, J. Phys. {\bf A15}, 2879 (1982);
            H. Meirovitch and A. Lim, J. Phys. Chem. {\bf 91}, 2544 (1989).
\bibitem{5} K. Kremer and K. Binder, Comp. Phys. Reports, {\bf 7}, 259 (1988);
            A. Baumgartner and K. Binder, {\it Application of Monte Carlo
            methods in Statistical Physics} (Springer, Berlin, 1984).
\bibitem{6} S. L. Narasimhan, P. S. R. Krishna, K. P. N. Murthy and M. Ramanadham,
            Phys. Rev. {\bf E65}, 010801(R) (2002).
\bibitem{7} I. Majid, N. Jan, A. Coniglio and H. E. Stanley,
            Phys. Rev. Lett. {\bf 52}, 1257 (1984).
\bibitem{8} S. L. Narasimhan, V. Sridhar, P. S. R. Krishna and K. P. N. Murthy,
            J. Phys. {\bf A} (submitted).
\bibitem{9} P. H. Poole, A. Coniglio, N. Jan and H. E. Stanley,
            Phys. Rev. {\bf B39}, 495 (1989).
\bibitem{10} V. Velikov, S. Borick and C. A. Angell, Science, {\bf 294}, 2335 (14 Dec.2001).

\end{references}
\end{document}